Superconductivity in the orthorhombic phase of thermoelectric $CsPb_xBi_{4-x}Te_6$ with $0.3 \leq x \leq 1.0$

R. X. Zhang,[1] H. X. Yang,[1,*] H. F. Tian,[1] G. F. Chen,[1,2] S. L. Wu,[1] L. L. Wei,[1] J. Q. Li [1,2,*]

[1] Beijing National Laboratory for Condensed Matter Physics, Chinese Academy of Sciences, Beijing 100190, China

[2] Collaborative Innovation Center of Quantum Matter, Beijing 100084, China

Experimental measurements clearly reveal the presence of bulk superconductivity in the $CsPb_xBi_{4-x}Te_6$ ($0.3 \leq x \leq 1.0$) materials, i.e. the first member of the thermoelectric series of $Cs[Pb_mBi_3Te_{5+m}]$, these materials have the layered orthorhombic structure containing infinite anionic $[PbBi_3Te_6]^-$ slabs separated with $Cs^+$ cations. Temperature dependences of electrical resistivity, magnetic susceptibility, and specific heat have consistently demonstrated that the superconducting transition in $CsPb_{0.3}Bi_{3.7}Te_6$ occurs at Tc=3.1K, with a superconducting volume fraction close to 100% at 1.8K. Structural study using aberration-corrected STEM/TEM reveals a rich variety of microstructural phenomena in correlation with the Pb-ordering and chemical inhomogeneity. The superconducting material $CsPb_{0.3}Bi_{3.7}Te_6$ with the highest Tc shows a clear ordered structure with a modulation wave vector of $\mathbf{q} \approx a^*/2 + c^*/1.35$ on the a-c plane. Our study evidently demonstrates that superconductivity deriving upon doping of narrow-gap semiconductor is a viable approach for exploration of novel superconductors.

PACS number: 74.70.-b, 74.62. Bf, 74.62.-c, 61.05.J-

In the past decades, the search for novel superconductors, especially unconventional superconductors has been considered as an important topic from both fundamental science and developments of technological applications. [1-3] Recent report of the occurrence of superconductivity at 4.4 K in p-type doped thermoelectric material $CsBi_4Te_6$ indicates that narrow-gap semiconductors can be a promising platform for exploding new superconducting materials.[4] However, it is noted that the reported $CsBi_4Te_6$ samples generally have relatively low superconducting fractions, e.g. a typical superconducting volume fraction about 14% for $Cs_{0.96}Bi_4Te_6$ and 20% for $CsBi_{4.1}Te_{5.9}$.[4] Our recent study has been focused on the $CsPbBi_3Te_6$ which is a typical number of a homologous series $Cs[Pb_mBi_3Te_{5+m}]$ with m=1.[5-7] These materials have orthorhombic layered structures and have been intensively studied in the past decades due to their excellent thermoelectric property. In structural point of view, both $CsPbBi_3Te_6$ and $CsBi_4Te_6$ possess a "$M_4Te_6$" type of framework, which can be viewed as a fragment excised from NaCl-type structure along the [011] direction with a thickness of four {NaCl} monolayers.[4-7] The $CsBi_4Te_6$ phase contains formally $Bi^{2+}$ ions, and distinct infinite $[Bi_4Te_6]^-$ rods are arranged side-by-side and linked with Bi-Bi bonds to make slabs, the $Cs^+$ ions are intercalated among the slabs.[4,8,9] Substitution of $Pb^{2+}$ for the $Bi^{2+}$ ions results in the orthorhombic $CsPbBi_3Te_6$ phase, and the $[PbBi_3Te_6]^-$ blocks evolve into the infinite slabs as shown in Fig. 1(a). In this paper, we will report on our discovery of superconductivity in the orthorhombic $CsPb_xB_{4-x}Te_6$ with the highest Tc = 3.1K at

x=0.3. Moreover, we performed a systematic aberration-corrected TEM/STEM study on $CsPb_xBi_{4-x}Te_6$ with $0.3 \leq x \leq 1$, the microstructure features and relevant ordered states corresponding to different Pb-orders are carefully analyzed. The superconductivity with highest Tc has been observed in $CsPb_{0.3}Bi_{3.7}Te_6$ in which a remarkable structural modulation associated with Pb-ordering can be written as **q**≈a*/2 + c*/1.35.

The $CsPb_xBi_{4-x}Te_6$ polycrystalline samples were prepared in a two-step solid reaction method: $Bi_2Te_3$, Pb, Te, and small excess (~1%) of Cs were firstly placed in a small alumina crucible, and then sealed in a silica tube. The tube was preheated at 200°C for 10h, then hold at 700°C for 40h, and then cooled inside the furnace to room temperature. The obtained products were then heated at 700°C for 40h, and air-quenched to room temperature. Microstructure and chemical composition were analyzed on a S-4800 scanning electron microscope (SEM). Specimens for transmission electron microscopy (TEM) observations were prepared by gently crushing the polycrystalline materials into fine fragments, which were then supported by a copper grid coated with a thin carbon film; TEM and STEM observations were performed in the JEOL ARM200F equipped with double aberration correctors and operated at 200kV. Under this condition, the spatial resolution is about 0.08 nm. Powder X-ray diffraction is performed on a Bruker AXS D8 advance diffractometer. The heat capacity measurement was performed by PPMS commercial device (Quantum Design) in the temperature range 1.8−8K and in magnetic fields B=0 and 1T. The temperature dependence of electrical resistivity ρ(T) was measured down to 1.8 K by a standard four-point probe method on a Quantum Design Physical Property Measurement System (PPMS). The temperature dependence of VSM magnetic susceptibility measurements were performed by using Quantum Design SQUID magnetometer at a magnetic field of 10 Oe.

X-ray diffraction measurements were firstly used to characterize the structure of $CsPb_xBi_{4-x}Te_6$ samples, Fig. 1(b) shows the experimental data for samples with nominal compositions of x=0, 0.1, 0.2, 0.3, 0.4, 0.5, 0.6, 0.7, 0.8, 0.9 and 1.0, respectively. It is recognizable that the crystal structure changes evidently with the increase of Pb doping content. The sample with x=0 has a C2/m structure, and the x=0.1, 0.2 sample shows certain complex structural features due to the coexistence of the monoclinic C2/m structure and an orthorhombic structure (Cmcm). Samples with x ranging from 0.3 to 1 can be well indexed by the orthorhombic structure and no impurity phases are observed. These facts suggest that the Pb substitution for Bi is of benefit for the stability of the orthorhombic phase. The structural refinement for the superconducting samples shows that the lattice parameters increase progressively with the increase of the Pb content. For instant, the $CsPb_{0.3}Bi_{3.7}Te_6$ (x=0.3) has the lattice parameters of a = 6.327(8) Å, b =28.648(4) Å, and c = 4.359(5) Å , and the lattice parameters of $CsPbBi_3Te_6$ (x=1.0) are a = 6.345(4) Å, b =28.680(8) Å, c =4.377(3) Å. Figure 1(c) shows a SEM image illustrating the morphological features of $CsPb_{0.3}Bi_{3.7}Te_6$ crystal grains, the direction of crystal growth is along the c-axis direction. Actually, the superconducting materials $CsPb_xBi_{4-x}Te_6$ often show very similar morphological features for x ranging from 0.3 to 1.0 as observed in our experiments.

Figure 2(a) shows the temperature-dependent magnetization for a series of $CsPb_xBi_{4-x}Te_6$ samples as measured in zero field-cooled (ZFC) modes. It is recognizable that all the samples show noticeable diamagnetic signal, the onset of the transition decrease slowly with the increase of the Pb content; especially nearly perfect diamagnetism was observed in $CsPb_{0.3}Bi_{3.7}Te_6$ confirming the bulk superconductivity in this layered system. The zero-field-cooled (ZFC) and field-cooled (FC) susceptibility data indicate a significant amount of vortices are pinned in the sample of $CsPb_{0.3}Bi_{3.7}Te_6$ as shown in the inset of Fig. 2(a). In general, the superconducting $CsPb_xBi_{4-x}Te_6$ always shows a bad metallic behavior and the resistivity decrease slowly with lowering temperature, similar to the low carrier density superconductors.[10-12] Figure 2(b) shows the typical temperature dependence of the resistivity of $CsPb_{0.3}Bi_{3.7}Te_6$. Above Tc, the value of resistivity is about 7 mΩ cm, which is an order of magnitude larger than that of typical metals. The onset of the sharp drop in resistivity occurs at 3.1 K and reaches zero at 2.3 K, as shown in the inset of Fig. 2(b). Furthermore, our investigations also show that the superconducting transition in resistivity for $CsPb_{0.3}Bi_{3.7}Te_6$ can be visibly suppressed under applied magnetic fields. The upper critical field is estimated from the field-dependent transition temperatures as 1.5 T using the Ginzburg−Landau (GL) law [13] and 1.8 T using the Wer-thamer−Helfand−Hohenberg (WHH) theory. [14] This data is much lower than the results as reported for $CsBi_4Te_6$, which shows very high critical field up to 9.7(5),[4] suggesting that the superconducting properties could be quiet different for these two compounds (see Supplementary Information, Fig. S1).

It is remarkable that our measurements of specific heat for $CsPb_{0.3}Bi_{3.7}Te_6$ demonstrate clear changes at the superconducting transition. Figure 2(c) shows the experimental data of specific heat divided by temperature (C/T) as plotted as a function of $T^2$. Normal-state specific heat was evaluated by suppressing the superconducting phase with a magnetic field of 1 T. In present case, we have fitted the normal-state specific-heat data below 7K using the formula of $C(T) = \gamma T+\beta_1 T^3+ \beta_2 T^5$, so we can obtain the relevant coefficients: $\gamma=8.75$ mJ•mol$^{-1}$•K$^{-2}$, $\beta_1=10.11$ mJ•mol$^{-1}$•K$^{-4}$, and $\beta_2=0.16$ mJ•mol$^{-1}$•K$^{-6}$. The Debye temperature $\Theta_D = [(12/5)NR\pi^4/\beta]^{1/3}$ is calculated to be about 128.4 K, where N and R are the number of atoms per formula unit and the gas constant, respectively. For the electronic specific heat $C_{el}$ in the superconducting state, a clear specific-heat jump, $\Delta C_{el}$, appears just below 3.09 K as exhibited in Fig. 2(d). The midpoint for present transition is determined to be 2.92 K in good agreement with the temperature-dependent of electric resistivity ($Tc^{mid}$=2.9 K) as shown in Fig. 2(b). The $\Delta C_{el}/Tc$ value is estimated to be 11.51 mJ•K$^{-2}$•mol$^{-1}$, then the dimensionless specific-heat jump $\Delta C/(\gamma Tc)$ is 1.32, which is slightly below the BCS theoretical value of 1.43, suggesting a weak coupling scenario for this superconducting material.

We now proceed to discuss the microstructural properties as observed in the superconducting materials, indeed our TEM observation often reveal the presence of superstructure modulations in superconducting samples and its periodicity and orientation changes progressively with Pb-concentrations. In particular, an incommensurate modulation changes regularly with Pb-contents appearing in the a-c crystal plane. Figures 3(a)-(c) show three typical electron diffraction patterns taken from the superconducting samples of $CsPb_xBi_{4-x}Te_6$ with x=0.3, 0.5 and 1.0, respectively. It is remarkable the modulation wave vector can be roughly written as $\mathbf{q} \approx a^*/2 + (1-\delta(x))c^*$, in which the incommensurability $\delta(x)$ are found to change non-linearly with the Pb concentration, which can be written as $\delta(x) \approx 1/6$ for x=0.5 and 1/4 for x=0.3. This fact suggests that this ordered state could be fundamentally in correlation with the Pb-ordering in the $[PbBi_3Te_6]^-$ block. In order to direct observe the structural changes in association with this structural modulation, we have performed high-resolution TEM observations on a few well characterized superconducting samples, our experimental results demonstrated that, though the notable overlap of different elements often appear in the TEM image taken along [010] direction, we can also clearly see the atomic structure with regular contrast changes appearing on the a-c crystal plane. Figure 3(d) shows a type high-resolution TEM image obtained from x=1.0 sample, in which the structural modulation adopts an simple superstructure vector of $\mathbf{q} = a^*/2 + c^*$, so a clear superlattice can be evidently seen in the atomic image, and a well-defined double-spaced structure as indicated by arrows apparently exists along the c-axis direction. On the other hand, the x=0.3 superconductor (Tc=3.1K) adopts an incommensurate structure and the wave vector can be written as $\mathbf{q} \approx a^*/2 + c^*/1.35$, therefore, atomic structures in high-resolution TEM image often show up complex ordered state as shown the in Fig. 3(e). It is recognizable the mixture structures with the periods of $L_1=2d_{200}$ and $L_2=3d_{200}$ can clearly observed in the shown area.

In order to better understand the layered structural feature and the common defect structures in this new superconducting system, we have also performed extensive structural investigations on the $CsPb_{0.3}Bi_{3.7}Te_6$ superconducting samples using selected-area electron diffraction and aberration-corrected STEM. Figures 4(a) and (b) show the electron diffraction patterns taken respectively along the relevant [100] and [001] zone-axis directions from a $CsPb_{0.3}Bi_{3.7}Te_6$ sample. The main diffraction spots with relatively strong intensity can be well indexed by orthorhombic structure in consistence with the x-ray-diffraction results as mentioned in above context. It is also noted that the diffraction pattern in Fig. 4(b) contains a series week satellite spots at the systematic position of (1/2 0 0), actually, these satellite spots are arising from the component part of the incommensurate modulations for x=0.3 sample and analyzed in Fig. 3 in above context for $\mathbf{q} \approx a^*/2 + c^*/1.35$.

The better and clear atomic image for revealing the stacking slabs and atomic sheets is obtained along the [001] zone axis direction, Figure 4(c) shows an STEM image demonstrates the structural layers with the stacking sequence of $Cs^+$ and [PbBi$_3$Te$_6$]$^-$ slabs along the b-axis direction, this atomic structure is in good agreement with the structural model as specifically illustrated in Fig. 4(d). It is well known that the brightness for an atomic column in the STEM photograph is proportional to $Z^n$ (with 1<n<2, and Z is the element number of the examined atom),[15] therefore the relatively brighter dots in present image correspond to the projections of the Pb/Bi columns, the small bright dots are the atomic images of the Te columns, and the Cs atomic layers are sandwiched among the [PbBi$_3$Te$_6$]$^-$ blocks.

Actually, in additional to Pb-ordering, our STEM observations of CsPb$_{0.3}$Bi$_{3.7}$Te$_6$ offers a rich variety of structural defects. Such as two typical local defects as indicated by arrows at a few positions in Fig. 4(c), it is recognizable that these defects often appear nearby the Cs layers in the STEM images. It was noted in previous literatures, structural refinements revealed that the occupancy of the Cs sites is about 85% for CsPbBi$_3$Te$_6$ and the final formula was written with a Bi-rich composition as Cs$_{0.85}$Pb$_{0.85}$Bi$_{3.15}$Te$_6$ to maintain electro-neutrality.[5,6] Our analysis on the local structural distortion and contrast anomalies from the experimental images suggest that both type of defects are closely related to an irregular Pb substitution at M(2) sites or Pb/Te substitution at Cs sites, as a result, the Cs vacancy (or Pb/Te interstitial atoms) appear in these structural layers for local structural relaxation. According to our experimental results, the presence of nonstoichiometric ratio in present superconductors and essential complex microstructure features of the thermoelectric series of Cs[Pb$_m$Bi$_3$Te$_{5+m}$] could be the fundamental reasons for the existence of these defect structures. Figure 4(e) and 4(f) clearly show the atomic images with visible contrast changes for two local defects, in which local structural distortions are also clearly exhibited. Indeed, this contrast change can be properly interpreted by a Pb substitution for a Cs ion in the Fig. 4(e) and interstitial Te atoms appear around a Cs ion in Fig. 4(f) as indicated by arrows.

In summary, we have discovered superconductivity in the orthorhombic $CsPb_xBi_{4-x}Te_6$ materials, which is the first member of the thermoelectric series of $Cs[Pb_mBi_3Te_{5+m}]$. Measurements of electrical resistivity, magnetic susceptibility, and specific heat clearly show bulk superconductivity existing in $CsPb_xBi_{4-x}Te_6$ ($0.3 \leq x \leq 1.0$) materials, and the highest Tc=3.1K is found in $CsPb_xBi_{4-x}Te_6$ with x=0.3. It is also shown that the Pb contents and its ordered states have significant effects on the superconductivity in this material system. Structural analysis reveals that this layered superconducting system contains a rich variety of microstructural phenomena, for instance, an incommensurate modulation depending on the Pb concentration have been observed in all superconducting samples, and the superconducting material with the highest Tc=3.1K contains a visible structural order with a wave vector of **q**≈a*/2+ c*/1.35 as clearly demonstrated in the high-resolution TEM images. Our study demonstrates that superconductivity deriving upon doping of narrow-gap semiconductor is a viable approach for exploration of novel superconductor.

Acknowledgement

This work was supported by the National Basic Research Program of China 973 Program (Grant Nos. 2011CB921703, 2015CB921300, 2011CBA00101, 2010CB923002, 2012CB821404), the Natural Science Foundation of China (Grant Nos. 11474323, 51272277, 91221102,11190022, 11274368), and by "Strategic Priority Re-search Program (B)" of the Chinese Academy of Sciences (No. XDB07020300 and XDB07020200).

**Reference**

[1] H. Takahashi, K. Igawa, K. Arii, Y. Kamihara, M. Hirano, and H. Hosono, Nature **453**, 376(2008).

[2] P. Chandra, P. Coleman, and R. Flint, Nature **493**, 621(2013).

[3] S. Pathak, VB. Shenoy, M. Randeria, and N. Trivedi, Phys. Rev. Lett. **102**, 027002(2009).

[4] C. D. Malliakas, D. Y. Chung, H. Claus, and M. G. Kanatzidis, J. Am. Chem. Soc. **135**, 14540(2013).

[5] K. F. Hsu, D. Y. Chung, S. Lal, T. Hogan, and M. G. Kanatzidis, Mater. Res. Soc. Symp. **691**, 269(2001).

[6] K. F. Hsu, D. Y. Chung, S. Lal, A. Mrotzek, T. Kyratsi, T. Hogan, and M. G. Kanatzidis, J. Am. Chem. Soc. **124**, 2410(2002).

[7] K. F. Hsu, S. Lal, T. Hogan, and M. G. Kanatzidis, Chem. Commun. **13**, 1380(2002).

[8] D. Y. Chung, T. Hogan, P. Brazis, M. Rocci-Lane, C. Kannewurf, M. Bastea,; C. Uher, and M. G. Kanatzidis, Science **287**, 1024(2000).

[9] D. Y. Chung, T. P. Hogan, M. Rocci-Lane, P. Brazis, J. R. Ireland, C. R. Kannewurf, M. Bastea, C. Uher, and M. G. Kanatzidis, J. Am. Chem. Soc. **126**, 6414(2004).

[10] M. Jourdan, and H. Adrian, Physica C **388**, 509(2003).

[11] F. F. Tafti, T. Fujii, A. Juneau-Fecteau, S. R. de Cotret, N. Doiron-Leyraud, A. Asamitsu, and L. Taillefer, Phys. Rev. B **87**, 184504(2013).

[12] S. Ganesanpotti; T. Yajima, K. Nakano, Y. Nozaki, T. Yamamoto, C. Tassel, Y. Kobayashi, and H. Kageyama, J. Alloys Compd. **613**, 370(2014).

[13] V. L. Ginzburg, and L. D. Landau, Zh. Eksp. Teor. Fiz. **20**, 1064(1950).

[14] N. R. Werthamer, E. Helfand, and P. C. Hohenberg, Phys. Rev. 147, 295(1966).

[15] S. J. Pennycook, and L. A. Boatner, Nature **336**, 565(1988).

**Figure caption:**

FIG. 1. (a) Structural model of the orthorhombic $CsPb_xBi_{4-x}Te_6$ phase, illustrating the layered structural features, M(1) and M(2) are two crystallographically distinct metal sites, octahedrally coordinated with Te atoms. (b) Powder diffraction data for a series of $CsPb_xBi_{4-x}Te_6$ samples, exhibiting a clear transition from a monoclinic to an orthorhombic structure. (c) A SEM image shows the microstructure of a typical superconducting sample, showing the needle-like morphology. The scale bar is 100 μm.

FIG. 2. (a) Experimental measurements of physical properties for superconducting samples. (a) Zero field cooling temperature-dependent magnetic susceptibility for $CsPb_xBi_{4-x}Te_6$. Inset shows the ZFC and FC data of $CsPb_{0.3}Bi_{3.7}Te_6$. (b) Temperature-dependent resistivity for $CsPb_{0.3}Bi_{3.7}Te_6$ at zero field. (c) and (d) specific heat data for $CsPb_{0.3}Bi_{3.7}Te_6$, demonstrating the presence of bulk superconductivity in present system.

FIG. 3. (a-c) Electron diffraction patterns taken from [010] zone-axis direction for samples with x=0.3, 0.5 and 1.0, respectively, illustrating the progressive changes of the structural modulation q with the increase of Pb-content. (d) High-resolution TEM image for x=1 sample showing the double-spaced structural feature along the c-axis direction. (e) High-resolution TEM image for x=0.3 superconductor, showing the mixed ordered states in correlation with the incommensurate modulation. The scale bar in both images is 1nm.

FIG. 4. Electron diffraction patterns of $CsPb_{0.3}Bi_{3.7}Te_6$ superconductor taken along (a) the [100], (b) the [001] zone-axis directions, these weak superstructure reflections actually are arising from component of the modulation as shown in figure 3(a)

$q \approx a^*/2 + c^*/1.35$. (c) Aberration-corrected-STEM image taken along the [001] zone axis direction, demonstrating clearly the layered structure and typical defect structures. The scale bar is 3nm. (d) An enlarged image come from the square-enclosed regions in (c), illustrating the layered structural features. Two typical defect structures, as indicated by arrows in (c), often appear in superconducting samples and illustrated in the enlarged image of (e) and (f). A visible atomic contrast change in (e) indicated by an arrow can be interpreted by a Pb substitution for Cs; additional bright dots in (f) are interpreted as interstitial Te atoms around a Cs ion.

**FIG. 1.**

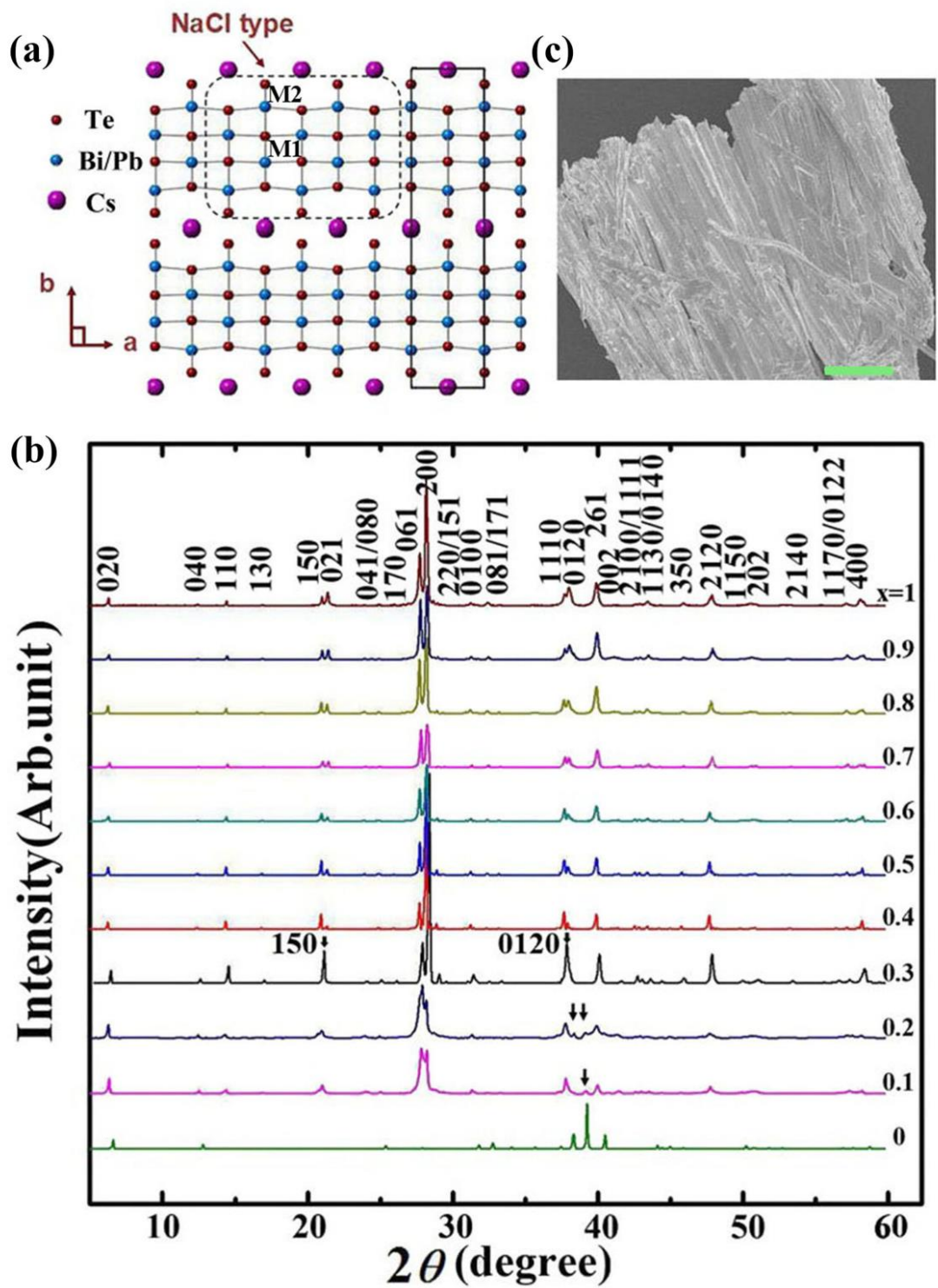

**FIG. 2.**

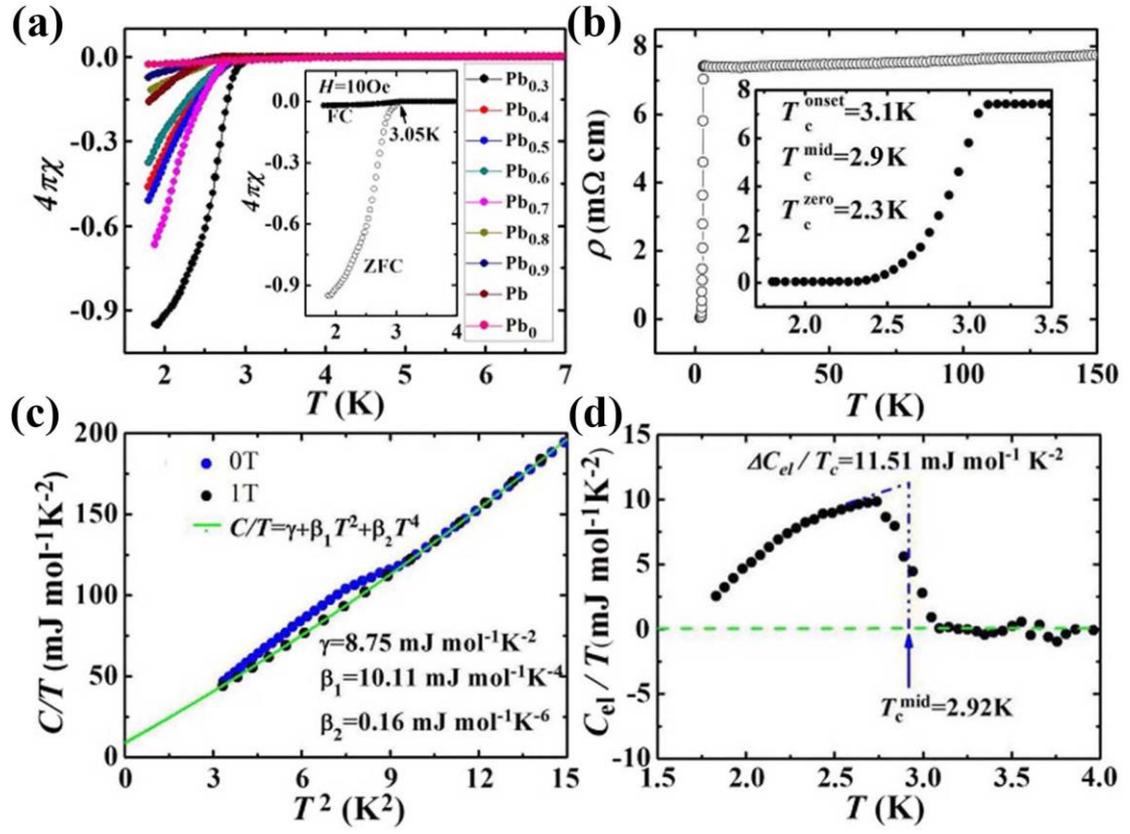

**FIG. 3.**

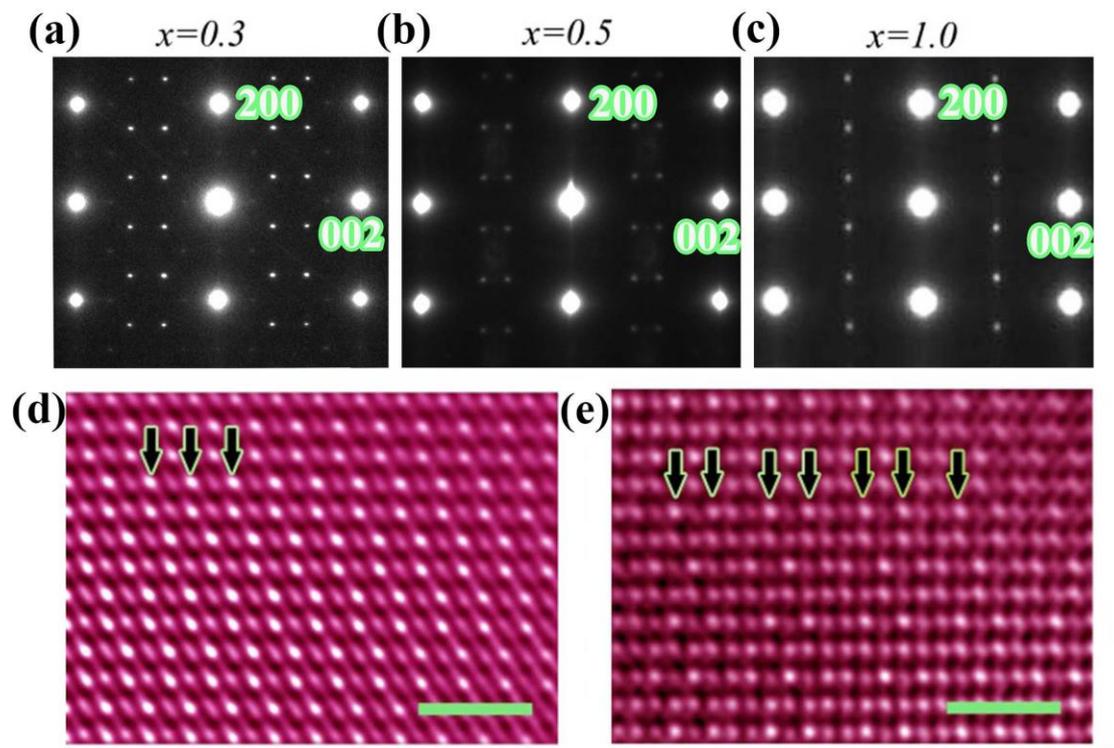

**FIG. 4.**

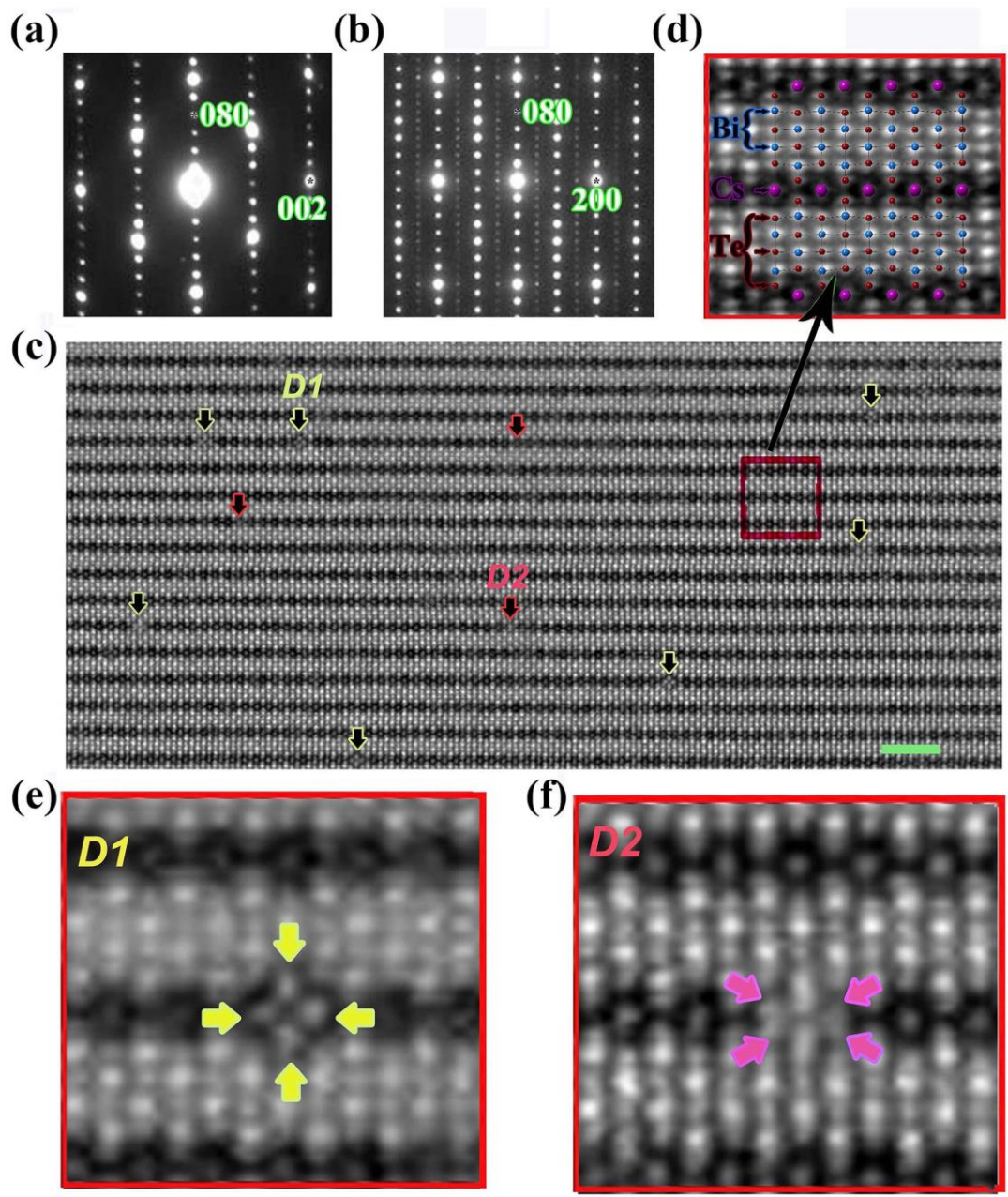

# Supplementary information

**1. Temperature-dependent resistivity at different field for CsPb$_{0.3}$Bi$_{3.7}$Te$_6$**

Our investigations show that the superconducting transition in resistivity for CsPb$_{0.3}$Bi$_{3.7}$Te$_6$ can be visibly suppressed under applied magnetic fields. The upper critical field is estimated from the field-dependent transition temperatures as 1.5 T using the Ginzburg−Landau (GL) law [1] and 1.8 T using the Wer-thamer−Helfand−Hohenberg (WHH) theory. [2] This data is much lower than the results as reported for CsBi$_4$Te$_6$, which shows very high critical field up to 9.7(5), [3] suggesting that the superconducting properties could be quiet different for these two compounds.

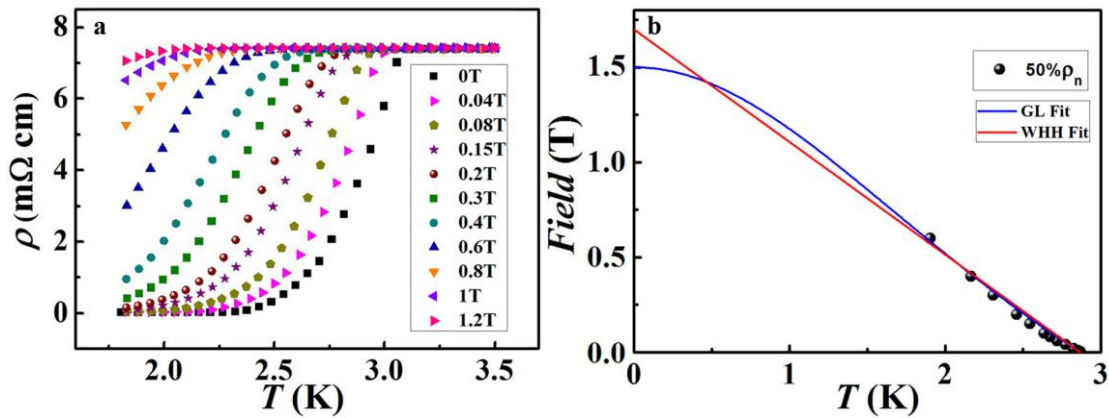

**Figure S1** (a) Temperature-dependent resistivity at different field for CsPb$_{0.3}$Bi$_{3.7}$Te$_6$, showing the systematic suppression of the superconducting transition. (b) The upper critical field curve obtained from the field-dependent transition temperatures. extraction of the upper critical field using the midpoint (50% criterion), and estimation of the upper citical field by Ginzburg−Landau (GL) law and Werthamer−Helfand−Hohenberg (WHH) theory gave data points up to1.5T and 1.8T respectively.

Reference:

[1] V. L. Ginzburg, L. D. Landau, Zh. Eksp. Teor. Fiz. **20**, 1064(1950).

[2] N. R. Werthamer, E. Helfand, P. C. Hohenberg, Phys. Rev. 147, 295(1966).

[3] C. D. Malliakas, D. Y. Chung, H. Claus, M. G. Kanatzidis, J. Am. Chem. Soc. **135**, 14540(2013).